\documentclass[reprint, aps,prd, preprintnumbers,groupedaddress,nofootinbib]{revtex4-1}

\pdfoutput=1
\usepackage{graphicx}
\usepackage{latexsym}
\usepackage{amsfonts}
\usepackage{amssymb}
\usepackage{amsmath}
\usepackage{slashed}
\usepackage{feynmp}
\usepackage{hyperref}
\usepackage{url}
\usepackage{color}
\usepackage{cancel}
\usepackage{multirow,array}

\usepackage{dcolumn}% Align table columns on decimal point
\usepackage{bm}% bold math

\def\bdrs{\text{BDRS}}

\newcommand\one{\leavevmode\hbox{\small1\normalsize\kern-.33em1}}

\newcommand{\gev}{{\ensuremath\rm GeV}}

\def\slashchar#1{\setbox0=\hbox{$#1$}           % set a box for #1
   \dimen0=\wd0                                 % and get its size
   \setbox1=\hbox{/} \dimen1=\wd1               % get size of /
   \ifdim\dimen0>\dimen1                        % #1 is bigger
      \rlap{\hbox to \dimen0{\hfil/\hfil}}      % so center / in box
      #1                                        % and print #1
   \else                                        % / is bigger
      \rlap{\hbox to \dimen1{\hfil$#1$\hfil}}   % so center #1
      /                                         % and print /
   \fi}

\newcommand{\be}{\begin{eqnarray*}}
\newcommand{\ee}{\end{eqnarray*}}

\newcommand{\bee}{\begin{eqnarray}}
\newcommand{\eee}{\end{eqnarray}}
\newcommand{\beeq}{\begin{equation}}
\newcommand{\eeeq}{\end{equation}}

\begin{document}

\title{Role of the $Z$ polarization in the $H \to b\bar{b}$ measurement}

\author{Dorival Gon\c{c}alves} 
\email{dorival@pitt.edu}
\affiliation{PITT PACC, Department of Physics and Astronomy, University of Pittsburgh, 3941 O'Hara St., Pittsburgh, PA 15260, USA}
\author{Junya Nakamura}
\email{junya.nakamura@itp.uni-tuebingen.de}
\affiliation{Institut f\"ur Theoretische Physik, Universit\"at T\"ubingen, 72076 T\"ubingen, Germany}

\preprint{PITT-PACC-1810}

%%%%%%%%%%%%%%%%%%%%%%%%%%%%%%%%%%%%%%%%%%%%%%%%%%
\begin{abstract}
It is widely known that the $ZH$ production channel provides a promising probe to the $H \to b\bar{b}$ 
decay at the LHC, when the Higgs and the $Z$ bosons  are highly boosted along the transverse direction.
We show how information on the $Z$ boson polarization -- that is disregarded in the current 
LHC analyses -- can relevantly improve the  signal from background discrimination, helping us to observe 
and achieve a larger precision for the $H \to b\bar{b}$ measurement  in the $Z(\ell\ell)H$ channel. 
\end{abstract}


\maketitle

%%%%%%%%%%%%%%%%%%%%%%%%%%%%%%%%%%%%%%%%%%%%%%%%%%
%\noindent
\section{Introduction}
The decay of the Standard Model Higgs boson to a bottom quark pair ($b\bar{b}$) has the largest branching
ratio among all Higgs decay modes, approximately 58\%~\cite{Heinemeyer:2013tqa}.  Measuring this decay is therefore crucial, not only 
for determining the coupling to bottom quarks, but also for constraining the total Higgs
boson decay width under reasonably general assumptions~\cite{Lafaye:2009vr,Corbett:2015ksa}. 

The largest sensitivity to $H \to b\bar{b}$ can be gained from the boosted $VH$ production, where 
$V$ denotes a $W$ or $Z$ boson that is boosted along the transverse direction; their leptonic decays result in
clean signatures which  can be efficiently  triggered on, while vetoing most of the multi-jet backgrounds~\cite{Butterworth:2008iy, 
Butterworth:2015bya}. 
Very recently ATLAS and CMS have reported observed significances of $4.9\sigma$ and $4.8\sigma$ for the
 $VH(b\bar{b})$ channel, and  $5.4\sigma$ and $5.6\sigma$ for the $H \to b\bar{b}$ decay, respectively, 
 in combinations with other Higgs production modes with their Run 1 and Run 2 data~\cite{Aaboud:2018zhk, Sirunyan:2018kst}. 
 Although the $VH(b\bar{b})$ is already very close to a $5\sigma$ significance, each channel 
 of the $VH(b\bar{b})$ process, categorized by the number of observed charged leptons, has a  considerably smaller significance. 
 For instance, the observed  significance for the two-lepton channel is $3.4 \sigma$ $(1.9 \sigma)$ for ATLAS (CMS) with the $13$~TeV 
 and $79.8~\mathrm{fb}^{-1}_{}$ ($41.3~\mathrm{fb}^{-1}_{}$) of  data~\cite{Aaboud:2018zhk, Sirunyan:2018kst}. 
 The significances of all the three channels are comparable~\cite{atlas_hbb,cms_hbb,Aaboud:2018zhk, Sirunyan:2018kst}. Therefore, an improvement in each channel would be important for a larger precision gain in the $H\rightarrow b\bar{b}$ measurement.

When a  search for the two-lepton $Z(\ell^+\ell^-)H(b\bar{b})$ channel is performed at the Large Hadron Collider (LHC), the dominant 
background after signal extraction procedures is $Zb\bar{b}$~\cite{atlas_hbb,cms_hbb,Aaboud:2018zhk, Sirunyan:2018kst}.  Although many theoretical 
efforts to identify the $H \to b\bar{b}$ decay have thrived since the pioneering work of Ref.~\cite{Butterworth:2008iy}, the $Z$ polarization 
has not been exploited to further discriminate the signal from the $Zb\bar{b}$ background\footnote{ATLAS and CMS~\cite{atlas_hbb,cms_hbb,Aaboud:2018zhk, Sirunyan:2018kst} use multivariate analyses to maximize the signal sensitivity.  In the two-lepton channel, the variables used for the multivariate 
analyses include only the two-lepton invariant mass $m_{\ell\ell}^{}$ as the information of the charged leptons. Hence,  the information on $Z$ boson polarization
is not used.}. In general, the $Z$ polarization can intrinsically differ from one process to another and manifests itself in the $Z\to \ell^+\ell^-$ decay 
angular distributions.

In this paper, we show that the $Z$ boson polarization has relevant information to  distinguish the signal from the dominant background, 
that is currently neglected~\cite{atlas_hbb,cms_hbb,Aaboud:2018zhk, Sirunyan:2018kst}. We   present a procedure to maximally exploit this information 
and estimate the possible sensitivity gains to the LHC analyses.

%%%%%%%%%%%%%%%%%%%%%%%%%%%%%%%%%%%%%%%%%%%%%%%%%%
%\vskip 0.1cm
%\no indent
\section{Approach}
The differential cross section including the $Z\to \ell^+\ell^-$ decay for both  $ZH$ and $Zb\bar{b}$ processes
 can be expanded in general as (we employ the notation of Ref.~\cite{Hagiwara:1984hi})
\begin{align}
& \frac{d\sigma}{dq_{\mathrm{T}}^2 dY d\cos{\Theta} d\cos{\theta}d\phi} =\nonumber \\
 & F_{1}^{} (1+\cos^2_{}{\theta} )
+ F_{2}^{} (1-3\cos^2_{}{\theta} )
+ F_{3}^{} \sin{2\theta} \cos{\phi}    \nonumber  \\
& + F_{4}^{} \sin^2_{}{\theta} \cos{2\phi} 
+ F_{5}^{} \cos{\theta}
+ F_{6}^{}  \sin{\theta} \cos{\phi}   \nonumber \\
& + F_{7}^{} \sin{\theta} \sin{\phi} 
+ F_{8}^{} \sin{2\theta} \sin{\phi} 
+ F_{9}^{} \sin^2_{}{\theta} \sin{2\phi}\,,
\label{differential}
\end{align}
where $q_{\mathrm{T}}^{}$ is the transverse momentum of the $Z$ boson in the laboratory frame, $Y$ is the rapidity of the $Z+H~(\mathrm{or\ }b\bar{b})$ system in the laboratory frame,  $\Theta$ is the polar angle of the $Z$ boson from the collision axis
in the $Z+H~(\mathrm{or\ }b\bar{b})$ center-of-mass frame, $\theta$ ($0 \le \theta \le \pi$) and $\phi$ ($0 \le \phi \le 2\pi$) are the polar and azimuthal 
angles  of the lepton ($\ell^-_{}$) in the $Z$ rest frame. We choose the coordinate system of the $Z$ rest frame following Collins 
and Soper (Collins-Soper frame)~\cite{Collins:1977iv}. This frame is well recognized and the angular coefficients for the Drell-Yan 
$Z$ production have been measured by ATLAS~\cite{Aad:2016izn, Aaboud:2017ffb} and CMS~\cite{Khachatryan:2015paa}. $F_i^{}$ are functions 
of only $q_{\mathrm{T}}^{}$, $Y$ and $\Theta$; integrations over other phase space variables are already performed.
After integrations over the lepton angles, only $F_1^{}$ remains;  $F_{1}^{}$ is directly related to the total 
cross section and determines the overall normalization of the $Z\to \ell^+\ell^-$ decay angular distributions. 
The eight functions $F_i^{}$ ($i=2$ to $9$) are described by polarization density matrices 
of the $Z$ boson.  

 Eq.~\ref{differential} opportunely simplifies for the signal and background processes under consideration.
First, we notice that the signal displays two relevant subprocesses: the quark-initiated Drell-Yan like $q\bar{q} \to ZH$ and
 the loop-induced gluon-fusion ${gg \to ZH}$. They are denoted by $ZH_{\mathrm{DY}}^{}$ and 
$ZH_{\mathrm{GF}}^{}$, respectively. 
Despite $ZH_{\mathrm{GF}}^{}$ being $\mathcal{O}(\alpha_s^2)$ suppressed, it results in important contributions at the boosted regime~\cite{Altenkamp:2012sx,
Englert:2013vua,Goncalves:2015mfa,Goncalves:2016bkl}. 
When CP non-conservation is neglected, the three coefficients $F_{7, 8, 9}^{}$ are always zero in tree-level calculations of any process~\cite{Hagiwara:1984hi}, because these are proportional to relative complex phases of scattering amplitudes. 
$ZH_{\mathrm{DY}}^{}$  can be analytically evaluated without difficulty at the leading-order (LO); we find that the scattering amplitudes for the $J_z^{}=0$ state of the $Z$ boson are zero, therefore  $F_{2,3,6}^{}$ are all zero. $F_5^{}$ is totally antisymmetric around $Y=0$, therefore it does not contribute after the integration over $Y$. Consequently, only $F_1^{}$ and $F_4^{}$ contribute after the $Y$ integration in $ZH_{\mathrm{DY}}^{}$. 
$ZH_{\mathrm{GF}}^{}$ receives constraints from CP conservation and Bose symmetry; as a result, $F_{5,6,8,9}^{}$ are zero at the LO. Although the coefficients $F_{3,7}$ are nonzero in $ZH_{\mathrm{GF}}^{}$, these are totally  antisymmetric around $\cos{\Theta}=0$,
thus do not contribute after the integration over $\cos{\Theta}$. Consequently, only $F_1^{}$, $F_2^{}$ and $F_4^{}$ contribute after the $\cos{\Theta}$ integration in $ZH_{\mathrm{GF}}^{}$. 
Estimation of the coefficients apart from $F_{7, 8, 9}^{}$ in the $Zb\bar{b}$ background process is not easy due to the large number of the scattering amplitudes. This process is part of the $\mathcal{O}(\alpha_s^2)$ correction to the Drell-Yan $Z$ production. Although the angular coefficients in this production have been calculated at $\mathcal{O}(\alpha_s^3)$ accuracy~\cite{Gauld:2017tww}, an exclusive calculation of those in the $Zb\bar{b}$ events does not exist in the literature to our knowledge. We have numerically found that only $F_1^{}$, $F_2^{}$ and $F_4^{}$ contribute after the integration over $Y$ and $\cos{\Theta}$, when the signal selections are imposed. The signal selections for the LO analysis are given later in Eq.~\ref{eq:eventcut1} and for the full hadron level study in Sec.~\ref{sec:results}.
To conclude, both for the signal and for the $Zb\bar{b}$ background, Eq.~(\ref{differential}) simplifies to 
\begin{align}
& \frac{d\sigma}{dq_{\mathrm{T}}^2 d\cos{\theta}d\phi} = \nonumber \\
&\widehat{F}_{1}^{} \bigl[ 1+\cos^2_{}{\theta} 
+ A_{2}^{} (1-3\cos^2_{}{\theta} )
 + A_{4}^{} \sin^2_{}{\theta} \cos{2\phi} \bigr],
 \label{differential-2}
\end{align}
where $A_{2} = \widehat{F}_2/\widehat{F}_1$ and $A_4 = \widehat{F}_4/\widehat{F}_1$. The hat above the coefficients implies that the integrations over $\cos{\Theta}$ and $Y$ are performed. For $ZH_{\mathrm{DY}}^{}$, we derive that
\begin{align}
A_4^{} = - \frac{q_{\mathrm{T}}^2}{ 2 m_Z^2 + q_{\mathrm{T}}^2},
\label{eq:A4}
\end{align}
which has a large negative value in the high $q_{\mathrm{T}}^{}$ region.
Eq.~(\ref{differential-2}) suggests that the angles $\theta$ and $\phi$
 can be defined in the restricted ranges $0 \le \theta \le \pi/2$ and $0 \le \phi \le \pi/2$ without losing any information. They can be obtained from
\begin{subequations}
\label{cosphi}
\begin{align}
|\cos{\theta}| & = \frac{2 \bigl| q^0_{} p_\ell^3 - q^3_{} p_\ell^0 \bigr|}{Q\sqrt{Q^2_{} + |\vec{q}_{\mathrm{T}}^{}|^2_{} }} \,, \label{costheta} \\
|\cos{\phi}| & = \frac{2}{\sin{\theta}}\frac{  \bigl| Q^2_{} \vec{p}_{\mathrm{T}\ell}^{}  \cdot  \vec{q}_{\mathrm{T}}^{} - |\vec{q}_{\mathrm{T}}^{}|^2_{} p_\ell^{} \cdot q \bigr|}{Q^2_{}|\vec{q}_{\mathrm{T}}^{}|\sqrt{Q^2_{} + |\vec{q}_{\mathrm{T}}^{}|^2_{}}} \,, 
\end{align}
\end{subequations}
where $q^{\mu}_{}=(q^0_{}, \vec{q}_{\mathrm{T}}^{}, q^3_{})$ and $p^{\mu}_{\ell}=(p^0_\ell, \vec{p}_{\mathrm{T}\ell }^{}, p^3_\ell)$ 
are four-momenta of the reconstructed $Z$ boson and one of the leptons, respectively, in the laboratory frame and $Q$ is the reconstructed $Z$ invariant 
mass ($Q=m_{\ell\ell}^{}$). We stress that $p_\ell^{\mu}$ can be the momentum  of either $\ell^-_{}$ or  $\ell^+_{}$ ({\it i.e.} either gives the same $\theta$
and $\phi$ values).
This is simply because interchanging $\ell^-_{}$ and $\ell^+_{}$ corresponds to $\theta \to \pi - \theta$  and $\phi \to \phi + \pi$ ({\it i.e.} $\cos{\theta} \to - \cos{\theta}$ and $\cos{\phi} \to - \cos{\phi}$). Practically speaking, we do not need to distinguish $\ell^-_{}$ and $\ell^+_{}$. This feature is particularly important for ${Z \to e^-_{} e^+_{}}$ in which case the charge misidentification rate is not negligible~\cite{atlas_hbb}. 

To evaluate the two coefficients $A_2$ and $A_4$, we simulate the signal and the  background at the LO with {\sc MadGraph5\_aMC@NLO}~\cite{Alwall:2014hca, Hirschi:2015iia}, using  the NNPDF2.3~\cite{Ball:2012cx}  parton distribution functions.  The $Z(\ell\ell)b\bar{b}$ background
sample accounts for the interference with $\gamma^{*}(\ell\ell) b\bar{b}$.  The events are required to pass the following  selections:
\begin{align}
&75~\gev < m_{\ell\ell}^{} < 105~\gev,\ \ 115~\gev < m_{bb} < 135~\gev,\nonumber \\
&p_{\mathrm{Tb}} > 25\ \mathrm{GeV},\ \ 
 |y_b| < 2.5,\ \ 0.3 < \Delta R_{bb} < 1.2 \,,
 \label{eq:eventcut1}
\end{align}
where $\Delta R_{bb}$ is a cone radius between the two b-quarks.
In Tab.~\ref{table:asymmetries}, we display the results in two $q_{\mathrm{T}}^{}$ regions. The
statistical uncertainty for the last digit is shown in parentheses. We find that both $ZH_{\mathrm{DY}}^{}$ 
and $ZH_{\mathrm{GF}}^{}$ present very distinct values $A_{2, 4}^{}$ from the $Zb\bar{b}$ background process. It is also observed that the $A_{4}^{}$ values in $ZH_{\mathrm{DY}}^{}$ are consistent with the analytic formula in Eq.~(\ref{eq:A4}). These differences clearly appear in the $(\cos\theta$, $\phi)$ distribution.
%-----------------
 \begin{table}[th!] 
 \vspace{0.3cm}
 \begin{ruledtabular}
 \begin{tabular}{rrrr}
       & $ZH_{\mathrm{DY}}^{}$ & $ZH_{\mathrm{GF}}^{}$ & $Zb\bar{b}$ \\
\colrule
$A_2^{}(q_{\mathrm{T}}^{}>200~\gev)$ & $0.001(1)$  & $0.026(1)$  & $0.470(1)$   \\
$A_2^{}(q_{\mathrm{T}}^{}>400~\gev)$ & $-0.002(3)$  & $0.052(8)$  & $0.498(4)$   \\
$A_4^{}(q_{\mathrm{T}}^{}>200~\gev)$ & $-0.825(2)$ & $-0.972(2)$ & $0.447(2)$   \\
$A_4^{}(q_{\mathrm{T}}^{}>400~\gev)$ & $-0.947(5)$ & $-0.92(1)$ & $0.462(8)$   
 \end{tabular}
 \end{ruledtabular}
  \caption{Normalized angular coefficients $A_2^{}$ and $A_4^{}$ in two regions of $q_{\mathrm{T}}^{}$ at the LO, after the selection in Eq.~\ref{eq:eventcut1}. The
statistical uncertainty for the last digit is shown in the parentheses. \label{table:asymmetries}}
 \end{table}
%-----------------

%-----------------
\begin{figure}[bh!]
\includegraphics[scale=0.7]{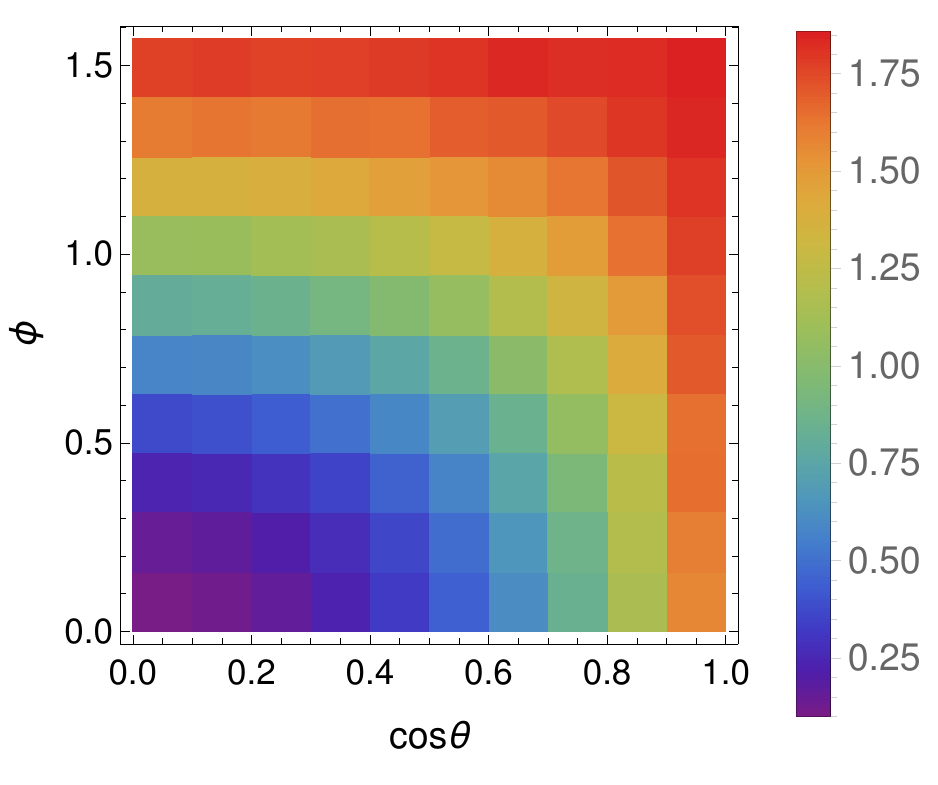}
\caption{Ratio of the normalized $(\cos{\theta},\phi)$ distribution at the LO for the $ZH_{\mathrm{DY}}^{}$ process to that for the $Zb\bar{b}$ background process, imposing the cuts in Eq.~(\ref{eq:eventcut1}) and $q_{\mathrm{T}}^{}>200$ GeV. \label{figure:csangles}}
\end{figure}
%-----------------

In Fig.~\ref{figure:csangles}, we show the ratio of the normalized $(\cos{\theta},\phi)$ profile for the $ZH_{\mathrm{DY}}^{}$ process to that for the $Zb\bar{b}$ process, imposing $q_{\mathrm{T}}^{}>200$ GeV. The information on the differences in $A_2^{}$ and $A_4^{}$ between the signal and the background completely appears in this restricted (i.e. $0 \le \theta \le \pi/2$ and $0 \le \phi \le \pi/2$) two-dimensional ${(\cos\theta, \phi)}$ distribution. 
For example, it is observed that signal events are more distributed at $\phi \sim \pi/2$ and $Zb\bar{b}$ more at ${\phi \sim 0}$, as the consequence of the large difference in $A_4^{}$ between the $ZH_{\mathrm{DY}}^{}$ and $Zb\bar{b}$ processes.  Therefore, the $Z\rightarrow \ell^+\ell^-$ decay angular distributions can be useful in distinguishing the signal from the background.\medskip

So far, minimum selections in the lepton transverse momentum $p_{\mathrm{T}\ell}^{}$ are not considered. 
Defining the lepton angles in the Collins-Soper frame has a great advantage: $p_{\mathrm{T}\ell}^{}$ has a simple expression in terms of these angles.
 In the laboratory frame, in which the $x$-axis is chosen along 
$\vec{q}_{\mathrm{T}}^{}$, the vectorial transverse momenta of the harder ($\ell_{1}^{}$) and softer $(\ell_2^{})$ leptons are given by
\begin{widetext}
\begin{subequations}
\label{eq:leptonpT}
\begin{align}
\vec{p}_{\mathrm{T\ell_{1}^{}(\ell_{2}^{})}}^{} = \frac{1}{2} \Bigl( q_{\mathrm{T}}^{} \pm \sqrt{ Q^2_{} + q_{\mathrm{T}}^2 } \sin{\theta} \cos{\phi}, \pm Q\sin{\theta} \sin{\phi} \Bigr)\,.
\end{align}
Therefore, their absolute values are given by
\begin{align}
p_{\mathrm{T\ell_{1}^{}(\ell_{2}^{})}}^{} \equiv \bigl|\vec{p}_{\mathrm{T\ell_{1}(\ell_{2})}}^{} \bigr| = \frac{1}{2} \sqrt{ q_{\mathrm{T}}^2 + Q^2_{} \sin^2{\theta} + q_{\mathrm{T}}^2 \sin^2{\theta} \cos^2{\phi} \pm 2 q_{\mathrm{T}}^{} \sqrt{ Q^2_{} + q_{\mathrm{T}}^2 } \sin{\theta} \cos{\phi}  }\,, \label{eq:leptonpTmag}
\end{align} 
\end{subequations}
\end{widetext}
which are now independent of a choice of the $x$-axis in the laboratory frame. In the boosted kinematic region $Q^2_{}/q_{\mathrm{T}}^2 \ll 1$, we derive
\begin{align}
p_{\mathrm{T}\ell_{1}^{}(\ell_{2}^{})}^{}  = \frac{1}{2} q_{\mathrm{T}}^{} \bigl( 1 \pm \sin{\theta} \cos{\phi} + \mathcal{O}\bigl(Q^2_{}/q_{\mathrm{T}}^2\bigr) \bigr).\label{eq:leptonptlimit}
\end{align} 
This in fact shows that $Z$ polarization can largely affect the lepton $p_{\mathrm{T}}^{}$ distributions even if the $Z$ is highly boosted. 
Representative differences in $p_{\mathrm{T}\ell}^{}$ between the signal and the background can be revealed as follows. 
In the phase space where the signal is more concentrated $\phi \sim \pi/2$, we find
\begin{align}
p_{\mathrm{T\ell1}}^{} = p_{\mathrm{T\ell2}}^{} = \frac{1}{2} \sqrt{ q_{\mathrm{T}}^2 + Q^2_{} \sin^2{\theta}}. \label{eq:pTsignal} 
\end{align} 
While in the region where the background displays more events ${\phi \sim 0}$, we have
\begin{align}
p_{\mathrm{T\ell1(2)}}^{} &= \frac{1}{2} \Bigl| q_{\mathrm{T}}^{} \pm \sqrt{ Q^2_{} + q_{\mathrm{T}}^2} \sin{\theta} \Bigr|, \nonumber \\
p_{\mathrm{T\ell1}}^{} - p_{\mathrm{T\ell2}}^{} & = \mathrm{min}\Bigl\{ q_{\mathrm{T}}^{}, \sqrt{ Q^2_{} + q_{\mathrm{T}}^2} \sin{\theta} \Bigr\}.\label{eq:pTback}
\end{align}
%
%-----------------
\begin{figure}[t!]
\centering
\includegraphics[scale=0.48]{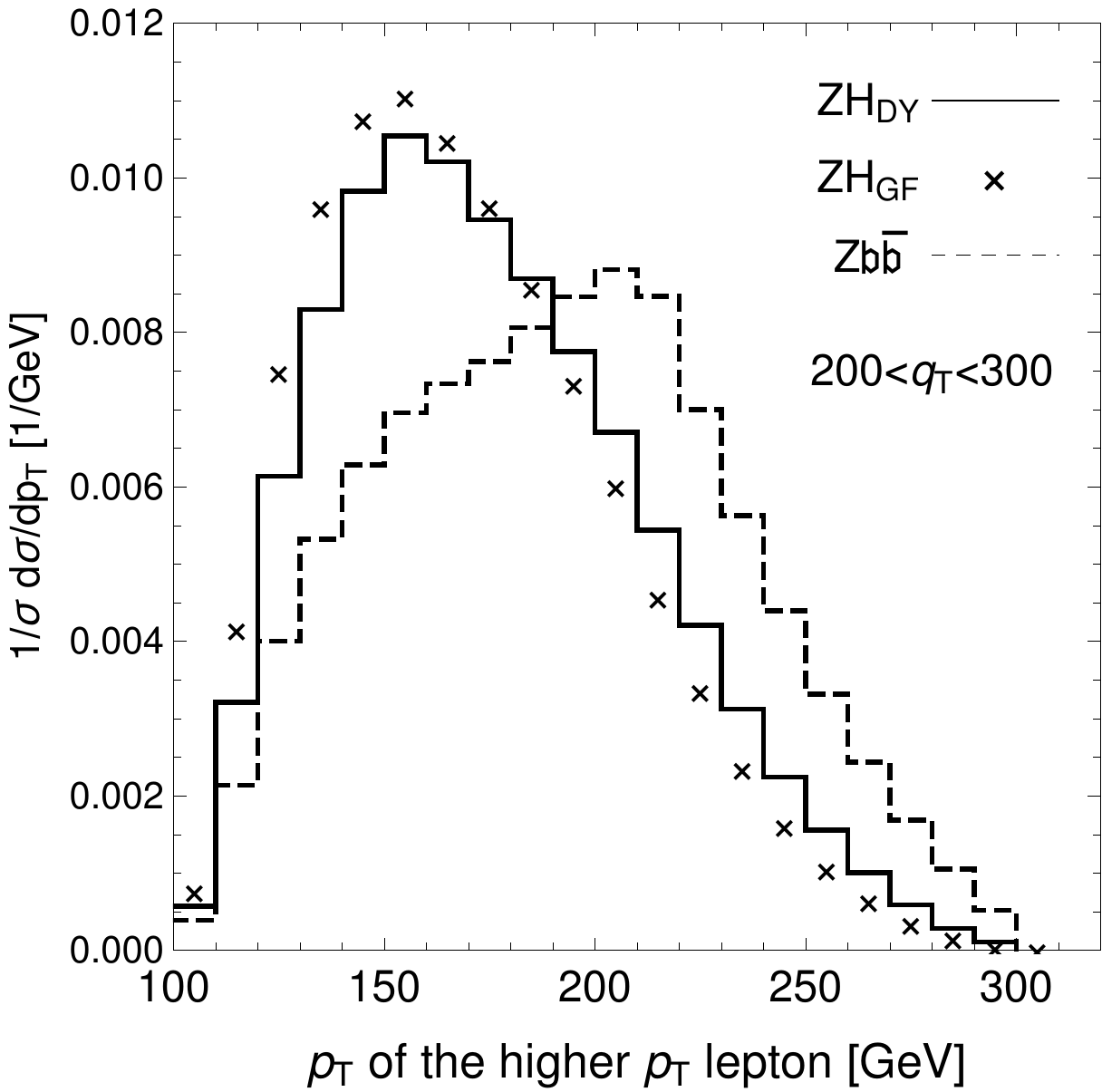}
\includegraphics[scale=0.48]{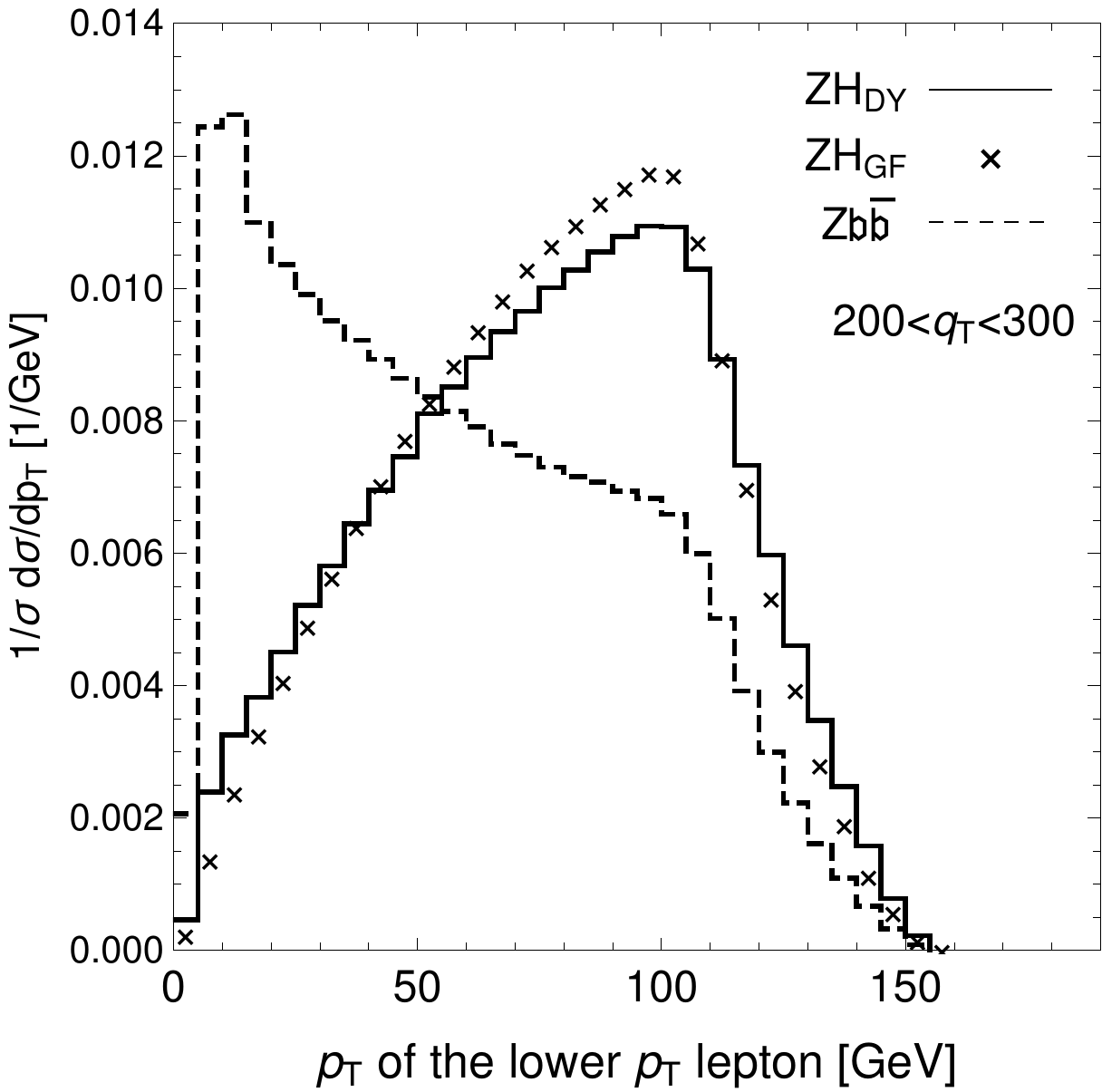}
\caption{\small Normalized $p_{\mathrm{T}}^{}$ distributions of the harder lepton  (left) and softer lepton (right)  for the 
$ZH_{\mathrm{DY}}^{}$ (solid curve), $ZH_{\mathrm{GF}}^{}$ ($\times$)  and $Zb\bar{b}$ (dashed curve) processes.
\label{figure:pT}}
\end{figure}
%-----------------
Eqs.~\ref{eq:pTsignal} and \ref{eq:pTback} tell us that
the higher (lower) $p_{\mathrm{T}}^{}$ lepton in the background is predicted to be harder (softer) than that in the signal. 
To illustrate this feature,  we show in Fig.~\ref{figure:pT} the normalized distributions of $p_{\mathrm{T\ell1(2)}}^{}$ for the
$ZH_{\mathrm{DY}}^{}$ (solid curve), $ZH_{\mathrm{GF}}^{}$ ($\times$) and $Zb\bar{b}$ (dashed curve) processes at
 the LO. The cuts in Eq.~\ref{eq:eventcut1}, ${200~\gev < q_{\mathrm{T}}^{} <300}$~GeV and a lepton rapidity 
cut $|y_\ell^{}|<2.5$ are imposed. The distributions roughly follow Eqs.~\ref{eq:pTsignal} and \ref{eq:pTback} which were
derived under the extreme conditions ({\it i.e.} $\phi=\pi/2$ and $\phi=0$). This observation confirms that the differences
in $p_{\mathrm{T\ell1(2)}}^{}$ distributions are consequences of the difference in the $Z$ polarization. It is, therefore, 
expected that a lepton $p_{\mathrm{T}}^{}$ selection can partially capture the difference in the $Z$ polarization ($A_4^{}$ in particular) 
and improve the sensitivity to the signal. 

We note in passing that the results in Eqs.~\ref{eq:pTsignal} and \ref{eq:pTback} and Fig.~\ref{figure:pT} 
illustrate the fundamental importance of taking account of  the $Z$ polarization in its decays in the Monte Carlo simulations, 
even without performing a tailored polarization analysis.

%%%%%%%%%%%%%%%%%%%%%%%%%%%%%%%%%%%%%%%%%%%%%%%%%%
%\vskip 0.1cm
%\noindent
\section{Results}\label{sec:results}
We now estimate the potential sensitive gains from the
 polarization information to a detailed hadron level ${pp\rightarrow Z(\ell\ell)H(b\bar{b})}$
LHC study. Our signal is characterized by two charged
leptons, $\ell=e$ or $\mu$, which reconstruct a boosted $Z$ boson recoiling against two $b$-jets. 
The major backgrounds are $Zb\bar{b}$, $t\bar{t}$+jets, and $ZZ$. 

We simulate our samples with Sherpa+OpenLoops~\cite{Gleisberg:2008ta,Cascioli:2011va,Denner:2016kdg}.
The $ZH_{\mathrm{DY}}$ and $ZH_{\mathrm{GF}}^{}$ signal and  $Zb\bar{b}$ and $ZZ$ background samples 
are merged at LO up to one extra jet emission via the CKKW algorithm~\cite{Catani:2001cc, Hoeche:2009rj}. 
The $ZH_{\mathrm{DY}}$, $Zb\bar{b}$ and $ZZ$ cross sections are normalized to the NLO rates obtained from Ref.~\cite{Goncalves:2015mfa}.
Finally, the $t\bar{t}$ is generated  at NLO with the MC@NLO algorithm~\cite{Frixione:2002ik,Hoeche:2011fd}.  Hadronization  and  
underlying event effects are taken into account in our simulation~\cite{Winter:2003tt}.

Pile-up is not simulated. The expected effects of pile-up relevant to our analysis are a degradation of the lepton isolation and  $b$ tagging performance; see e.g.~Sec. 6.4 of~\cite{Collaboration:2272264}. Pile-up affects both the signal and the dominant background $Zb\bar{b}$ equally, and leads to a lower sensitivity overall. Therefore, it would not alter the main conclusions of our study on the improvement in the signal significance gained from $Z$ polarization.

We follow the BDRS~\cite{Butterworth:2008iy} analysis for tagging the $H\to b\bar{b}$ as a well understood benchmark. However, we stress that our proposal uses only the lepton information, completely independent of how the $H\to b\bar{b}$ tagging is performed. 
We require two leptons, which have the same flavor and opposite-sign charges, with 
$|\eta_\ell |<2.5$ in the invariant mass range ${75~\gev<m_{\ell\ell}<105~\gev}$.
The $Z$ boson is required to have a large boost  $q_{\mathrm{T}}^{}\equiv p_{\mathrm{T}\ell\ell}^{} >200~\gev$. The hadronic 
activity is reclustered with the Cambridge-Aachen jet algorithm~\cite{Dokshitzer:1997in,Wobisch:1998wt,Cacciari:2011ma} with 
${R=1.2}$,  requiring at least one boosted $(p_{\mathrm{T}J}>200~\gev)$  and central $(|\eta_{J}|<2.5)$ fat-jet. This must be 
Higgs-tagged via the BDRS algorithm, demanding three sub-jets with the hardest two being $b$-tagged. Our study assumes a flat
70\% $b$-tagging efficiency and 1\% miss-tag rate. To further enhance the signal to background ratio, we demand the filtered Higgs
mass to be in the range  ${|m_{H}^{\bdrs}-m_H^{}|<10~\gev}$ with $m_H^{}=125$ GeV. The resulting event rate  is presented in Tab.~\ref{tab:cuts_analysis}, for which ${p_{\mathrm{T}\ell}>30}$~GeV is imposed. 

%-------------------------------------------------------
\begin{table}[h!]
\begin{tabular}{l  || c | c | c | c | c  }
 \multirow{1}{*}{} &
 \multicolumn{1}{c|}{$ZH_{\mathrm{DY}}^{}$}  &
 \multicolumn{1}{c|}{$ZH_{\mathrm{GF}}^{}$} &
  \multicolumn{1}{c|}{$Zb\bar{b}$} &
 \multicolumn{1}{c|}{$t\bar{t}$}  &
  \multicolumn{1}{c}{$ZZ$}   \\
  \hline 
{BDRS reconstruction} & \multirow{2}{*}{0.16} & \multirow{2}{*}{0.03} & \multirow{2}{*}{0.35} & \multirow{2}{*}{0.02} & \multirow{2}{*}{0.02} \\
$|m_{H}^{\bdrs}-m_H|<10~\gev$  &  &  &   &   &  \\
\end{tabular}
  \caption{Signal $ZH_{\mathrm{DY}}^{}$ and $ZH_{\mathrm{GF}}^{}$ and background $Zb\bar{b}$, 
  $t\bar{t}$,  and $ZZ$ rate after the BDRS analysis. Hadronization   and underlying event effects are taken into account. The rates
   are given in units of fb and take account of 70\%  $b$-tagging efficiency and 1\% misstag rate.}
\label{tab:cuts_analysis}
\end{table}
%-------------------------------------------------------

%-------
\begin{figure}[b!]
\includegraphics[scale=0.55]{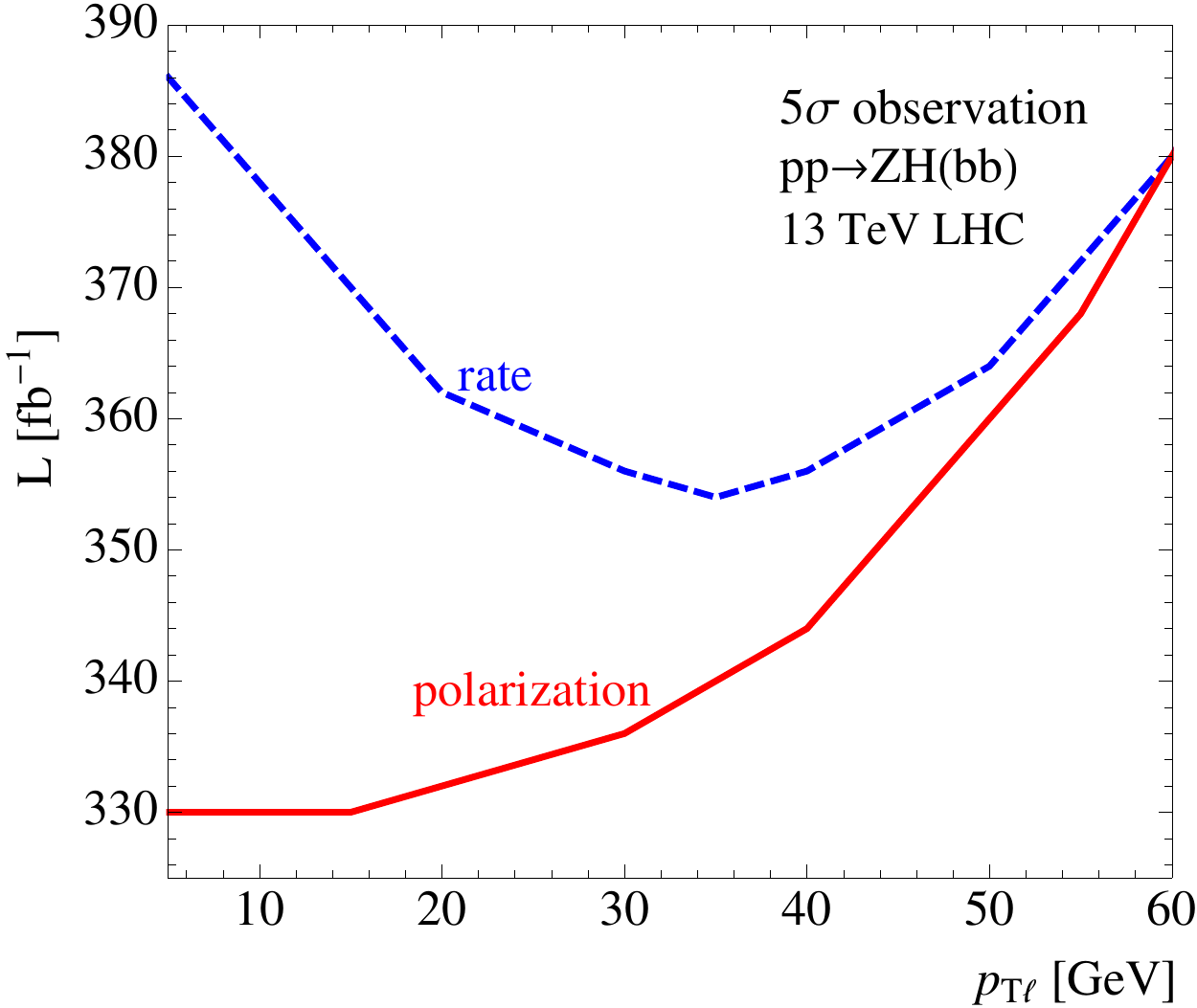}
\caption{\small Luminosities required for a $5\sigma$ observation of the two-lepton $pp\rightarrow Z(\ell\ell)H(bb)$
channel as functions of the lepton $p_{\mathrm{T}}^{}$ lower threshold. The solid curve takes the $Z$ polarization
 into account by performing a two dimensional binned log-likelihood analysis on the $(\cos{\theta},\phi)$ distribution, while
  the  dashed curve accounts only for the rate. 
\label{figure:luminosity}}
\end{figure}
%-------

To quantify  the importance of the $Z$ polarization, we perform a two dimensional
binned log-likelihood analysis based on the $(\cos{\theta},\phi)$ distribution with an
uniformly $5\times5$ binned grid, invoking the CL$_s$ method~\cite{Read:2002hq}.
In  Fig.~\ref{figure:luminosity}, we display the luminosities required for a $5\sigma$ observation as functions of the 
lower $p_{\mathrm{T}\ell}$  threshold. The solid curve represents the result of the polarization analysis, while the dashed curve is the 
result without the polarization study and thus purely rate based. 
The latter continuously improves from 385~$\mathrm{fb}^{-1}$ to 355~$\mathrm{fb}^{-1}$ by changing the $p_{\mathrm{T}\ell}^{}$ threshold from 5 to 35~GeV. This is precisely what we have expected as a consequence of the difference in the polarization; see the discussion below Eq.~\ref{eq:pTback}. It does not improve anymore above 35 GeV, because the statistics of the signal is also much depleted. 
On the other hand, the solid curve shows that the required luminosity monotonically increases with the $p_{\mathrm{T}\ell}^{}$ threshold.
This is because the polarization information (more practically the $(\cos{\theta},\phi)$ distribution) is disturbed by the $p_{\mathrm{T}\ell}^{}$ lower cut. This selection can capture the difference in polarization only partially, therefore it is never better than the explicit use of the polarization information. It is, therefore, suggested to define $p_{\mathrm{T}\ell}^{}$ threshold as small as possible and to exploit the difference in the $(\cos{\theta},\phi)$ distribution between the signal and the background. We stress that, because our proposal relies only on lepton reconstruction, it can be readily included in the current ATLAS and CMS studies. The current ATLAS (CMS) study~\cite{Aaboud:2018zhk, Sirunyan:2018kst} uses the $p_{\mathrm{T}\ell}^{}$ threshold 7~GeV (20~GeV), in which case the benefit of exploiting the $Z$ polarization is estimated to be $\sim15\%$ $(\sim10\%)$ in the required   luminosity for 5$\sigma$ observation.
%-------
\begin{figure}[t!]
\includegraphics[scale=0.55]{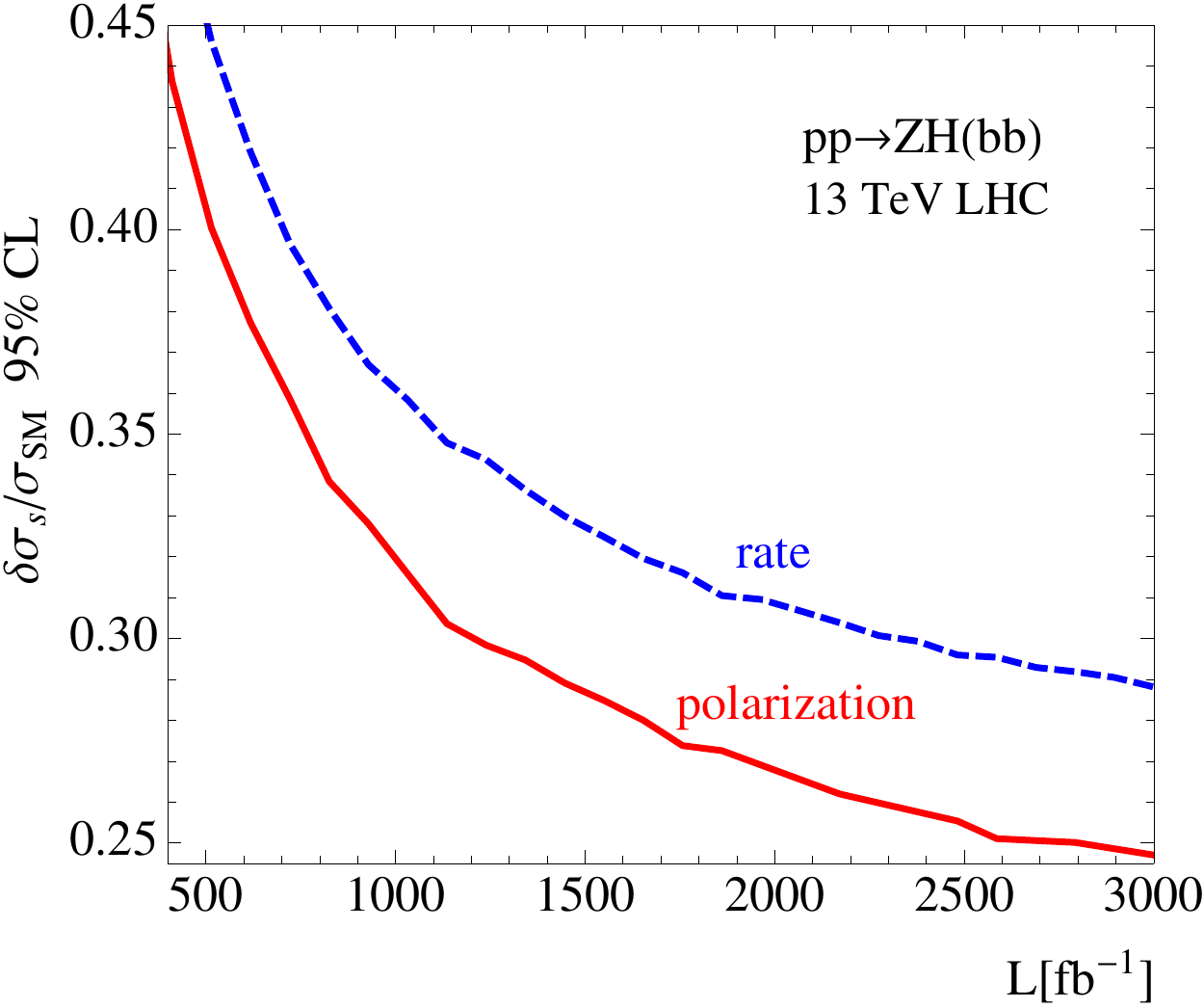}
\caption{\small 95\% confidence level bounds for separating an anomalous $Z(\ell\ell)H(b\bar{b})$ signal strength from the 
Standard Model prediction as a function of the LHC luminosity. The solid curve accounts for the $Z$ polarization
 via a two dimensional binned log-likelihood analysis on  $(\cos{\theta},\phi)$, while  the  dashed curve accounts only for the rate. 
 We assume a lepton $p_{\mathrm{T}}^{}$ threshold of $p_{\mathrm{T}}^{}>5$~GeV.
\label{figure:mu}}
\end{figure}
%-------

 In Fig.~\ref{figure:mu}, we show the 95\% confidence level bounds for separating an anomalous $Z(\ell\ell)H(b\bar{b})$  signal strength from
 the Standard Model prediction as a function of the 13~TeV LHC luminosity. We assume 5\% systematic uncertainties on the backgrounds, 
 which are modeled as nuisance parameters. The result with the polarization considerably enhances the precision on the signal strength determination and  makes the bounds less systematic limited  at high luminosities. This is a result of the larger $\mathcal{S}/\mathcal{B}$ for
several bins in the 2-dimensional phase space $(\cos\theta,\phi)$, see Fig.~\ref{figure:csangles}. While only 1.5~ab$^{-1}$ is required for
the polarization analysis to achieve  a precision of $29\%$ on the signal strength, if we disregard the polarization, one would demand the 
double of the data,  3~ab$^{-1}$, to achieve the same precision. 

%%%%%%%%%%%%%%%%%%%%%%%%%%%%%%%%%%%%%%%%%%%%%%%%%%
%\vskip 0.1cm
%\noindent
\section{Summary and discussion}
We have studied the potential of the $Z$ polarization to improve the sensitivity to the signal $pp\rightarrow Z(\ell\ell)H(b\bar{b})$.
At first, we have shown that the signal and the $Zb\bar{b}$ dominant background exhibit  different states of  $Z$ polarization, whose information completely appears as a large difference in the restricted two-dimensional $(\cos\theta,\phi)$ distribution, where $\cos\theta$ and $\phi$ parametrize the lepton momentum in the $Z$ rest frame. 
This difference can be partially captured by a suitable value for the $p_{\mathrm{T}\ell}^{}$ lower threshold, and fully taken into account by explicitly analyzing  the  $(\cos\theta,\phi)$ distribution. 
We have estimated the impact of these two approaches on the $5\sigma$ observation and that of the latter approach on $95\%$ CL upper bound on the uncertainty in the signal strength determination 
in the two-lepton $Z(\ell\ell)H(b\bar{b})$ channel, and found  relevant improvements. 
Since this proposal relies only on lepton reconstruction, displaying small experimental uncertainties, it can be promptly 
included in the current ATLAS and CMS studies.

In our study, we consider inclusive events in a high $q_{\mathrm{T}}^{}$ region ($q_{\mathrm{T}}^{}>200$ GeV), namely we do not make use of $F_i^{}$ dependences on $q_{\mathrm{T}}^{}$, $Y$ and $\cos{\Theta}$. The $q_{\mathrm{T}}^{}$ dependence is shown in Table~\ref{table:asymmetries}; the difference between $ZH_{\mathrm{DY}}^{}$ and $Zb\bar{b}$ is a little enhanced as $q_{\mathrm{T}}^{}$ becomes high. 
Some of the $Y$ and $\cos{\Theta}$ dependences are already described above Eq.~(\ref{differential-2}). In the $Z(\ell^+\ell^-)H(b\bar{b})$ channel, $Y$ and $\cos{\Theta}$ can be reconstructed for each event. It may be, therefore, possible, by using these additional information, to achieve a better signal sensitivity.

With our encouraging results, our approach could be applicable to other important channels, {\it e.g.}, {\it i)}~$Z(\ell\ell)H(\mathrm{invisible})$ and 
{\it ii)}~$W(l\nu)H(b\bar{b})$. We conclude with several comments on these two channels. {\it i)}~Needless to say, the $Z$ polarization is independent of how the Higgs boson decays; the $Z$ polarization in the $ZH(\mathrm{invisible})$ is the same as that in the $ZH(b\bar{b})$. The dominant background in this channel is $Z(\ell\ell)Z(\nu\nu)$. The $ZZ$ process shows almost the same $A_{2, 4}^{}$ values as the $Zb\bar{b}$ process. See Ref.~\cite{Goncalves:2018ptp} for more details.
{\it ii)} Despite of the neutrino in the final state, by assuming that the charged lepton and the neutrino construct a $W$ boson nominal mass, $|\cos{\theta}|$ and $|\cos{\phi}|$ in Eq.~(\ref{cosphi}) are still uniquely determined~\cite{Hagiwara:1984hi}. Since the $W^-_{}$ and $W^+_{}$ are always in the same state of polarization~\cite{Nakamura:2017ihk}, we can simply add $W^-_{}H$ events and $W^+_{}H$ events. The $WH$ and the irreducible background $Wb\bar{b}$ show the similar $A_{2, 4}^{}$ values as the $ZH_{\mathrm{DY}}^{}$ and the $Zb\bar{b}$, respectively. Details will be published elsewhere~\cite{Goncalves:2018}. 

%%%%%%%%%%%%%%%%%%%%%%%%%%%%%%%%%%%%%%%%%%%%%%%%%%
\begin{acknowledgments}
We thank C. Borschensky and S. Hasegawa for useful discussions. 
DG was funded by U.S. National Science Foundation under the grant PHY-1519175. JN appreciates the support from the Alexander von Humboldt Foundation.
\end{acknowledgments}\medskip

\bibliography{paper}

\end{document}